\documentclass{PoS}

\usepackage{amsmath}

\title{Chiral Condensate and Susceptibility of SU(2) $n_f=8$ Naive Staggered System}

\ShortTitle{Chiral Condensate and Susceptibility of SU(2) $n_f=8$ Naive Staggered System}

\author{\speaker{Issaku Kanamori}\\
        RIKEN Center for Computational Science, Kobe, Hyogo 650-0047, Japan\\
        E-mail: \email{kanamori-i@riken.jp}}

\author{C.-J. David Lin\\
Institute of Physics, National Chiao-Tung University, 1001 Ta-Hsueh Road, Hsinchu 30010, Taiwan\\
Centre for High Energy Physics, Chung-Yuan Christian University,
Chung-Li, 32032, Taiwan\\
        E-mail: \email{dlin@mail.nctu.edu.tw}}

\abstract{
The SU(2)  gauge theory with 8 fundamental fermions is studied using unimproved staggered regularization.  A phase transition or a crossover
at strong coupling, which can be a bulk transition.
By using chiral random matrix model we analyze the chiral condensate 
of this system.  We also report the chiral susceptibility and 
its volume dependence near the transition point.
}

\FullConference{37th International Symposium on Lattice Field Theory - Lattice2019\\
		16-22 June 2019\\
		Wuhan, China}

\begin{document}

\section{Introduction}

Identifying gauge theories in the conformal window
has been attracting a lot of attention, in the context of composite Higgs model
as well as purely theoretical interests.
One of the simplest candidates is SU(2) gauge theory with fundamental fermions.
According to the Helsinki group with their Wilson fermion simulations,  
the system is in the conformal window
with $n_f=6$ and $n_f=8$ flavors \cite{Leino:2017lpc,Leino:2017hgm}.
We also studied the $n_f=8$ system with naive staggered fermion 
and surveyed the bulk phase structure \cite{Huang:2014xwa,  Huang:2015vkr}.
In \cite{Huang:2015vkr}, we used chiral symplectic Random Matrix Theory (RMT
with the Dyson index $\beta_{\mathrm{D}}=4$) to analyze the chiral condensate of the system.  
In order to obtain the smallest eigenvalue distribution of the RMT, 
we used a finite matrix rank $N=400$ and used Hybrid Monte Carlo method.
We observed that the chiral symmetry is broken in the strong coupling side, 
while it is restored in the weak coupling side.
The Polyakov loop analysis suggested that the transition is weakly first order \cite{Huang:2014xwa},
but from the chiral condensate, the order was unclear.

Recently, one of the authors (I.K.) together with H.~Fuji and S.~M.~Nishigaki
has provided a new numerical tool to estimate individual eigenvalue distribution for $\beta_{\mathrm{D}}=4$ RMT \cite{Fuji:2019oby}.
In this talk,  we apply this new formula to SU(2) $n_f=8$
staggered system and analyze chiral symmetry breaking pattern.
Combining HMC estimation of the all the eigenvalue distribution of 
RMT with larger $N$ and the revised estimation of the chiral condensate,
we further estimate the chiral susceptibility, as from its volume dependence
we can argue the nature of the bulk phase transition.

In the next section, we briefly review the relation of eigenvalue spectra
between RMT and QCD(-like) theory.  Then we present the revised result
of the chiral condensate.
The RMT analysis of the chiral susceptibility is given 
in Sec.~\ref{sec:susceptibility}.
Section.~\ref{sec:conclusions} is the conclusion.

\section{Chiral Condensate}
\label{sec:condensate}

We fit the value of chiral condensate with RMT by using the following relation:
\begin{equation}
  \rho_{\mathrm{QCD}}(m;\lambda_i) 
= \rho_{\mathrm{RMT}}(\mu    = V\Sigma_{\mathrm{param}}m,
                    \zeta_i= V\Sigma_{\mathrm{param}}\lambda_i).
 \label{eq:equiv}
\end{equation}
In the simulation for the QCD-like gauge theory, we input the fermion mass, $m$, and the four-volume and extract the $i-$th smallest eigenvalue of the Dirac operator.  The distributions of these small eigenvalues can also be determined.
From the RMT, we have a mass parameter $\mu$, $i$-th smallest eigenvalue $\zeta_i$ and its distribution $\rho_{\mathrm{RMT}}$.
Equation~(\ref{eq:equiv}) tells us 
that the distributions of the eigenvalues are identical with a rescaling
by $V\Sigma_{\mathrm{param}}$ of the eigenvalues and the mass parameters.
Here, $\Sigma_{\mathrm{param}}$ is the chiral condensate of the QCD side. 
Note that the above relation holds in the broken phase of chiral symmetry and
for $\lambda_i$ smaller than (the correspondence of) the Thouless energy.
In such a situation, we can fit the value of $\Sigma_{\mathrm{param}}$.
If the fit does not work to obtain $\Sigma_{\mathrm{param}}$, it implies
the QCD side is in the symmetric phase.
Note that we can safely assume that the smallest $\lambda_i$ is smaller than the Thouless energy even the system is not in the $\epsilon$-regime.

Our lattice setting is the following.
The action is plaquette gauge action with unimproved staggered fermions.
The gauge group is SU(2) and
the number of fermions is $n_f=8$ in the fundamental representation, for which
no rooting trick is needed.
That is, we use two staggered flavors.
Our fermion mass analyzed here are $am=0.003, 0.005, 0.010$, and
the lattice volume is in $L^3\times T = 6^3 \times 6$ -- $16^3 \times 16$.
We use the periodic boundary condition in all directions 
for both gauge field and fermions.
The distribution of the eigenvalues depends on the topological charge $\nu$
so that we choose to work with $\nu=0$ for the RMT, and use gauge configurations in the same
topological sector in the fitting.
The number of fermions for RMT, 
that is the degeneracy of the eigenvalues,
is rather non-trivial.
As pointed out in~\cite{BerbenniBitsch:1998sy}, 
it is $2n_f$ for $n_f$ flavor system due to the double-fold degeneracy coming form the pseudo reality of SU(2) gauge group.
In addition, we have observed no eigenvalue degeneracy for the staggered taste
for our parameters with naive staggered fermion.
As a result, 
we compare our QCD-like simulation with RMT for the number of flavors
$n_f^{\mathrm{RMT}}=2n_f/4 = 4$.

\begin{figure}
 \center
 \includegraphics[width=0.45\linewidth]{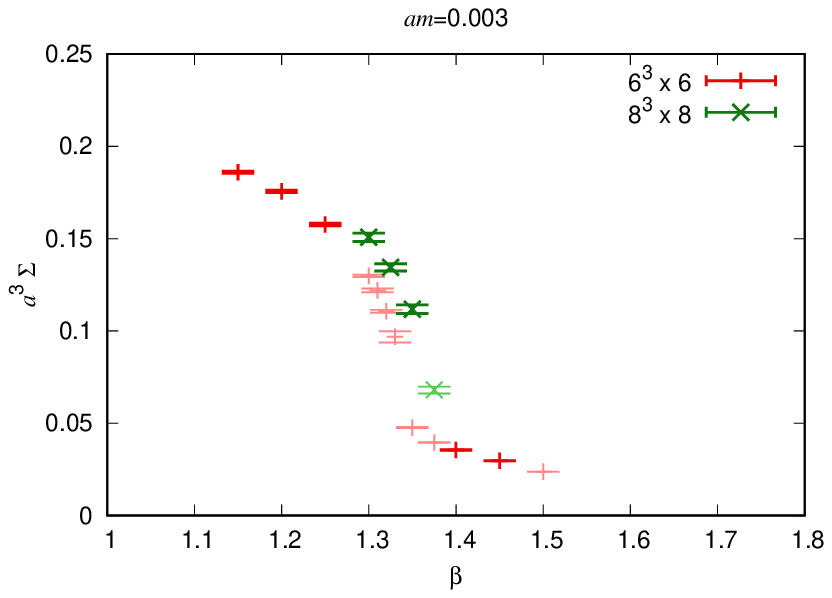}
 \hfil
 \includegraphics[width=0.45\linewidth]{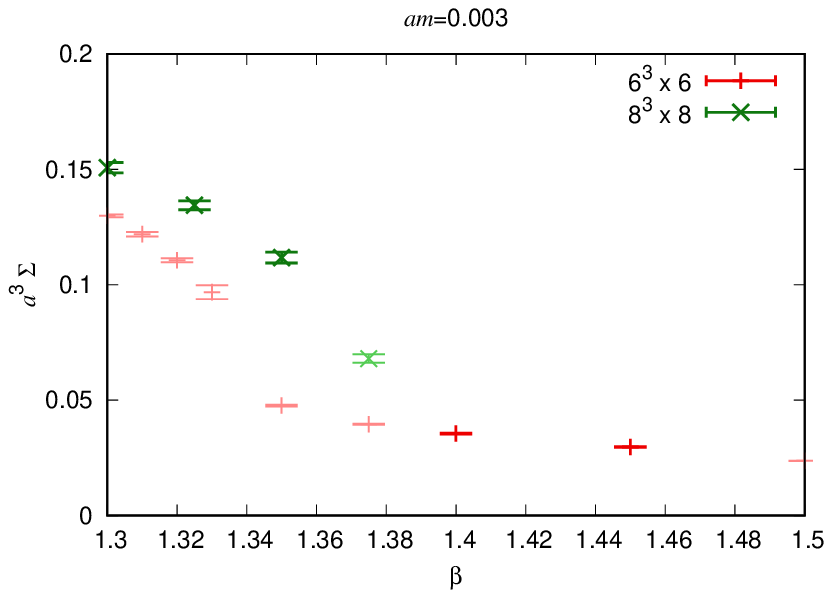}

 \center
 \includegraphics[width=0.45\linewidth]{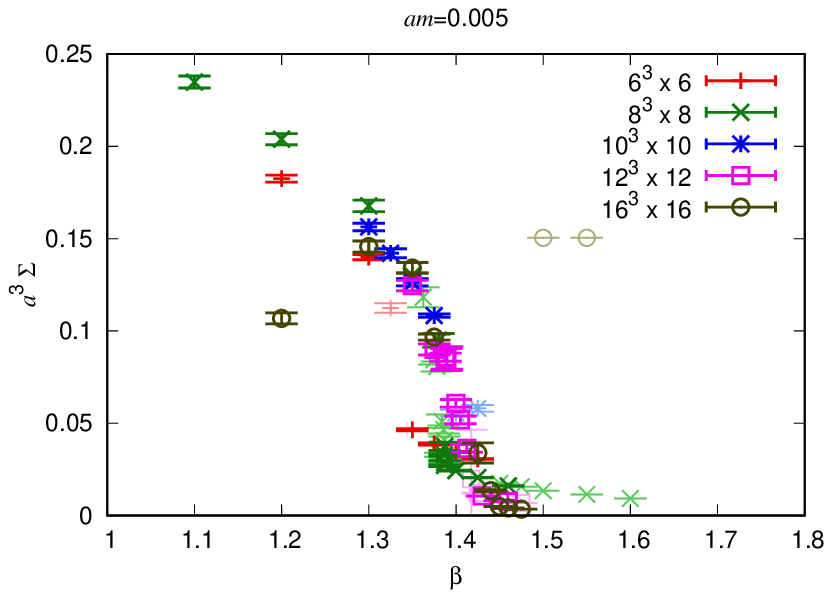}
 \hfil
 \includegraphics[width=0.45\linewidth]{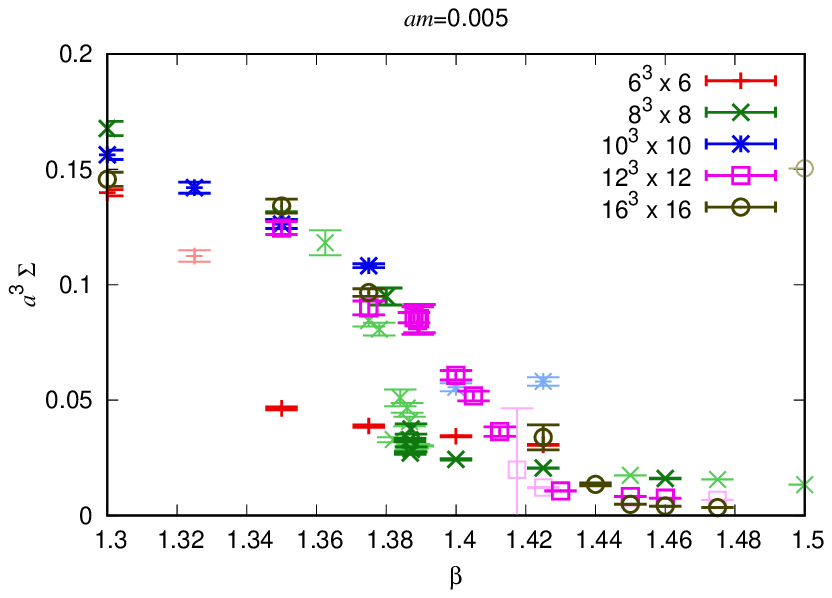}

 \center
 \includegraphics[width=0.45\linewidth]{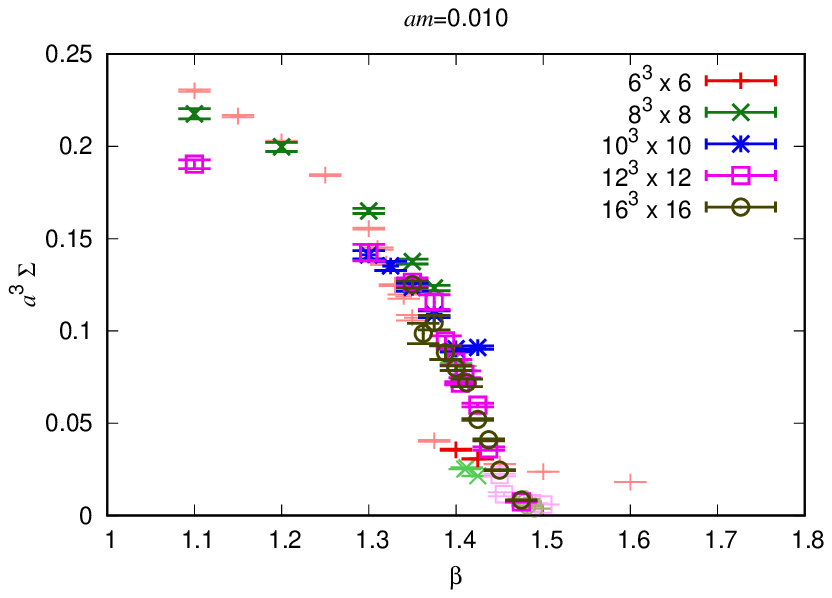}
 \hfil
 \includegraphics[width=0.45\linewidth]{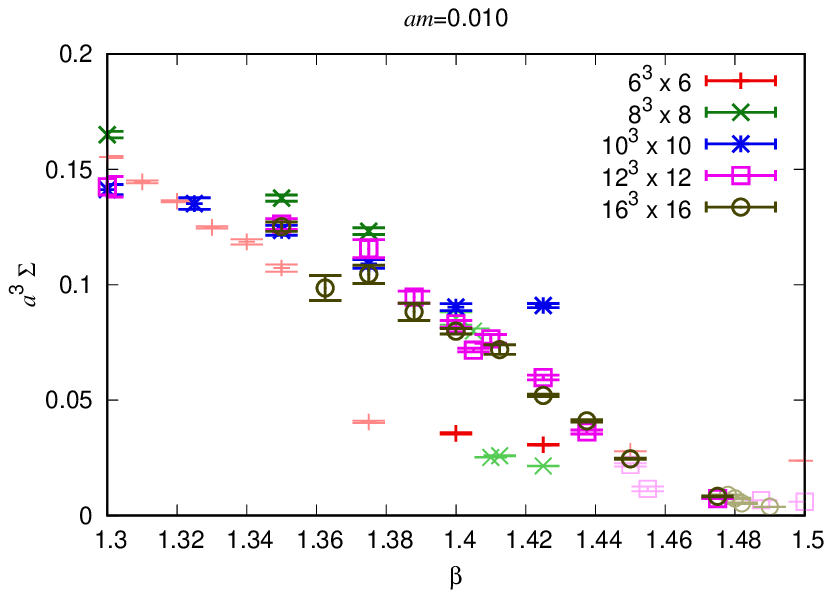}

\caption{Chiral condensates $\Sigma = \Sigma_{\mathrm{param}}$
in the lattice unit obtained by using RMT (left column) 
and the zoom-ups (right column).  
The fermion mass is $am=0.003$, $0.005$ and $0.010$, respectively, 
from the top panels to the bottom.
\label{fig:sigma}
}
\end{figure}

We fit the smallest eigenvalue from the lattice simulation by using
the RMT eigenvalue distribution obtained in  
\cite{Fuji:2019oby}\footnote{
In \cite{Fuji:2019oby}, it is shown that 
our previous estimate with $N=400$
\cite{Huang:2015vkr} has a sizable finite $N$ effects
to the estimation of the smallest eigenvalue distribution.
Compared with the error coming from lattice data in \cite{Huang:2015vkr}, 
however, it is the same order or smaller so that 
the systematic error coming from finite $N$ was under control.}.
We plot our fitted result of the chiral condensate in Fig.~\ref{fig:sigma}.
In the plot, light colored symbols are for fit with large chi squared
per degrees of freedom ($\chi^2/\mathrm{d.o.m}>1.5$).
The errors are obtained by jackknife analysis.
We observe finite chiral condensate at strong coupling,
i.e., small $\beta=4/g^2$. It disappears at around $\beta=1.4$--$1.5$.
As reported in \cite{Huang:2014xwa}, there is a bulk transition 
at the same $\beta$ value, where no four-volume dependence of the 
transition point appears in the plaquette variables.
We associate the transition between broken and symmetric phase 
of the chiral symmetry to the same bulk transition.
In order to access to the conformal window, we must perform the simulation 
in the symmetric phase, i.e., in the weak coupling side.
Although our previous result in~\cite{Huang:2015vkr} is based on a subset
of data used in this analysis, 
it is essentially the same as the revised result.

\section{Chiral Susceptibility}
\label{sec:susceptibility}

By noting that
 $
 Z_{\mathrm{QCD}}(m; \lambda_i)
  = Z_{\mathrm{RMT}}(\mu = V\Sigma_{\mathrm{param}}m; \zeta_i=V\Sigma_{\mathrm{param}}\lambda_i)
 $
in the $\epsilon$-regime, 
we have an expression of chiral susceptibility $\chi$:
\begin{align}
 \chi
 &= -\frac{1}{n_f}\frac{1}{V} \frac{\partial^2}{\partial^2 m}\ln Z_{\mathrm{QCD}} 
  = -\frac{1}{n_f}V\Sigma_{\mathrm{param}}\frac{\partial^2}{\partial \mu^2}
\ln Z_{\mathrm{RMT}} 
\\
 &= V\Sigma_{\mathrm{param}}^2\left\{
 \langle A(\mu) \rangle_{\mathrm{RMT}}
 + n_f \left(
 \langle B(\mu) B(\mu) \rangle_{\mathrm{RMT}}
 - \left( \langle  B(\mu) \rangle_{\mathrm{RMT}}\right)^2
 \right)
 \right\}\\
 &= \frac{\Sigma_{\mathrm{param}}}{m}{ \mu \left\{
 \langle A(\mu) \rangle_{\mathrm{RMT}}
 + n_f \left(
 \langle B(\mu) B(\mu) \rangle_{\mathrm{RMT}}
 - \left( \langle  B(\mu) \rangle_{\mathrm{RMT}}\right)^2
 \right)
 \right\}},
 \label{eq:susc}
\end{align}
where
$ A(\mu)
 = \sum_i \left[ \frac{2}{\zeta_i^2 +\mu^2}
 - \frac{4\mu^2}{\left(\zeta_i^2+\mu^2\right)^2}  \right]$
and 
$ B(\mu)
 = \sum_i \frac{2\mu}{\zeta_i^2 + \mu^2} $.
The expectation value $\langle \bullet \rangle_{\mathrm{RMT}}$ 
is that in the RMT.
The same formula was used for the Dyson index $\beta_{\mathrm{D}}=2$ system \cite{BerbenniBitsch:1999ti}.
Since we have the fitted result of the chiral condensate $\Sigma_{\mathrm{param}}$ and thus the value of $\mu$, we can calculate the chiral susceptibility $\chi$.
Although the parameter $\Sigma_{\mathrm{param}}$ gives the chiral condensate at the infinite volume, the partition function $Z_{\mathrm{QCD}}$ is the one at finite volume $V$.  
That is, the susceptibility $\chi$ obtained by eq.~(\ref{eq:susc}) has a volume dependence.

It is important to note that this formula is based on the equivalence of the whole partition function.  
That is, we must use the lattice data in the $\epsilon$-regime so that the value of $\mu$ must be order 1 or smaller.
This is different from fitting the chiral condensate, for which
we need only the information of the smallest eigenvalue, and gives 
a restriction on using eq.~(\ref{eq:susc}).
For the $\beta_{\mathrm{D}}=2$ quenched system,
it was shown in \cite{BerbenniBitsch:1999ti} that the formula leads to
a good agreement with the result obtained by lattice simulation 
at $\mu$ is $O(1)$ or smaller.
Since we need all the eigenvalues of the RMT, 
and it is not practically feasible to employ 
the new method in \cite{Fuji:2019oby} to achieve this, 
we resort to a numerical approach by formulating the problem as 
a one-dimensional field theory and simulating the theory using 
the Hybrid Monte Carlo (HMC) algorithm with the matrix rank $N=2000$. 
The details of the HMC on this system
is found in \cite{Fuji:2019oby}.

\begin{figure}
\center
\includegraphics[width=0.55\linewidth]{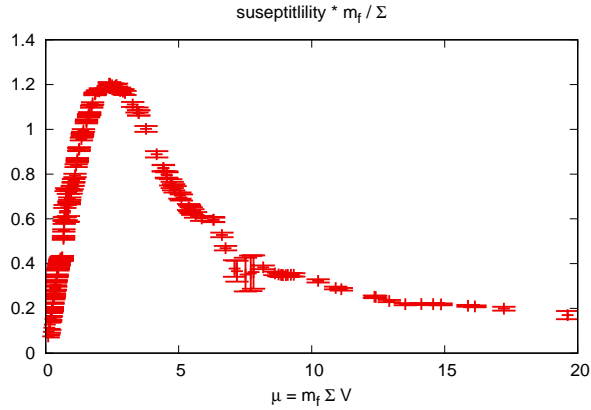} 
 \caption{The rescaled chiral susceptibility $m\chi/\Sigma=m\chi/\Sigma_{\mathrm{param}}$
obtained by using eq.~(\ref{eq:susc}).
}
 \label{fig:susc_mu_dep}

\end{figure}

Figure~\ref{fig:susc_mu_dep} shows the rescaled chiral susceptibility 
$m\chi/\Sigma_{\mathrm{param}}$ against 
$\mu=m\Sigma_{\mathrm{param}} V$. 
For a fixed value of fermion mass $m$ and 
the chiral condensate $\Sigma_{\mathrm{param}}$, 
which are equivalent to a fixed value of $\beta$ (see Fig.~\ref{fig:sigma}),
$\mu$ is proportional to the volume.  Therefor the linear raising of 
$m\chi/\Sigma_{\mathrm{param}}$ near the origin 
in Fig.~\ref{fig:susc_mu_dep} implies
a linear growth of the susceptibility in volume.  In this small $\mu$
region, however, no peak structure appears and thus it is not clear whether
this linear behavior is also true for the peak of the susceptibility or not.

\begin{figure}
\center
\includegraphics[width=0.49\linewidth]{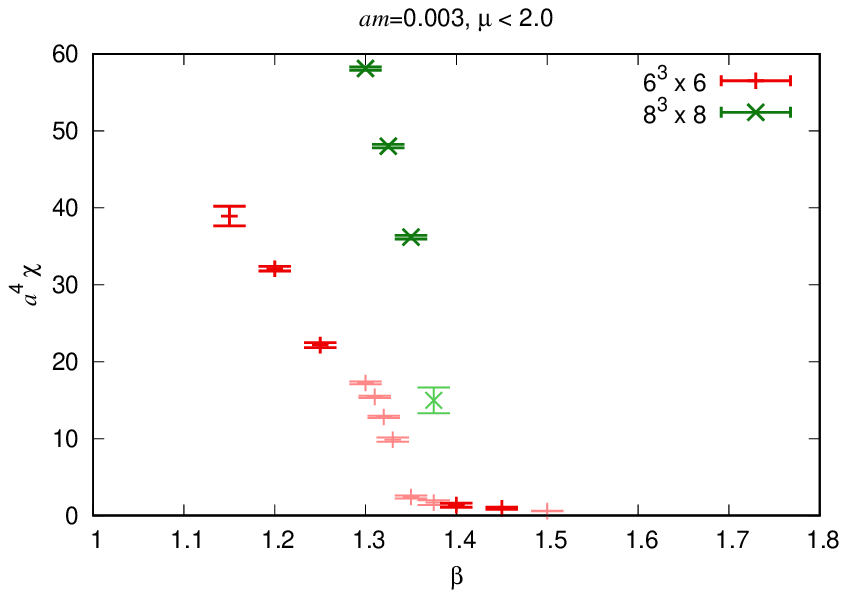}\\
\center
\includegraphics[width=0.49\linewidth]{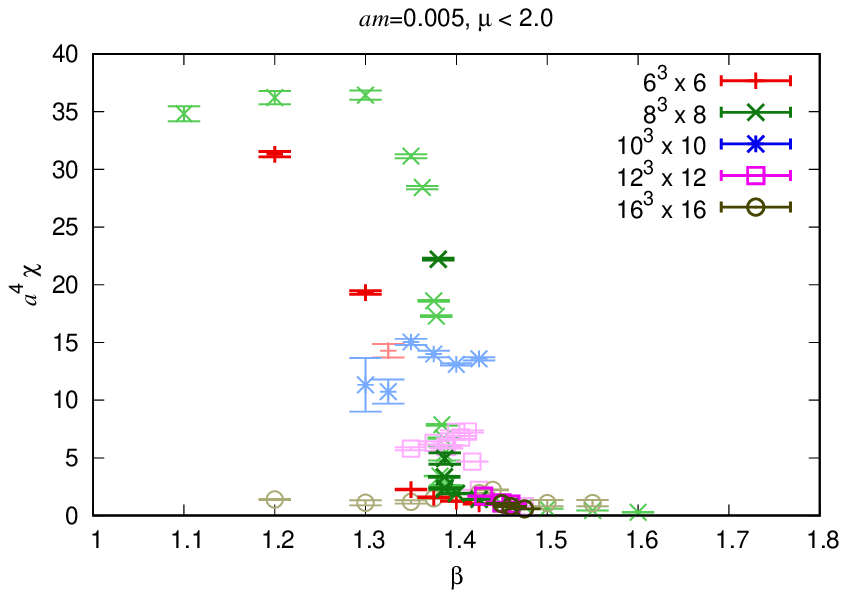}
\hfil
\includegraphics[width=0.49\linewidth]{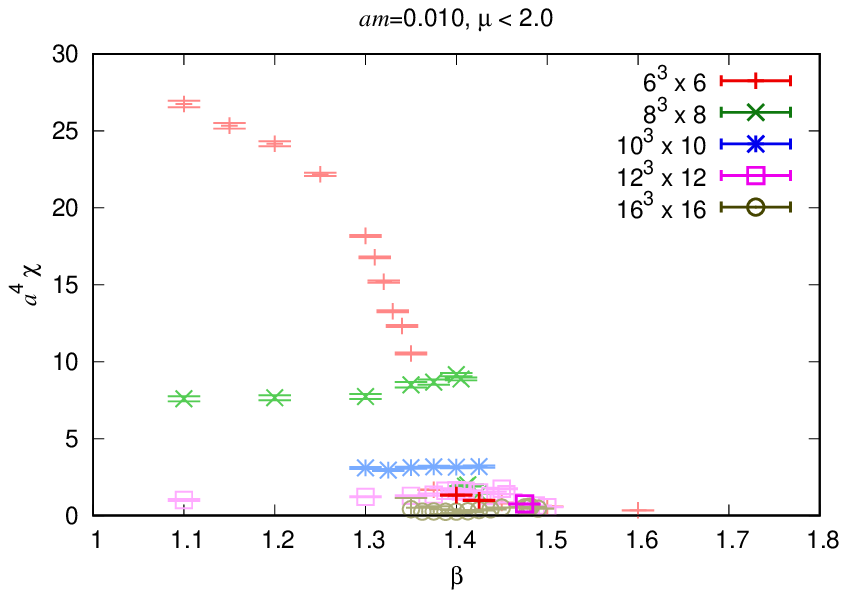}

 \caption{Chiral susceptibility.  Light color symbols correspond to 
$\mu \geq 2$ data, for which eq.~(\ref{eq:susc}) may not apply.}
 \label{fig:susc}
\end{figure}

In Fig.~\ref{fig:susc}, we plot the susceptibility obtained 
by using eq.~(\ref{eq:susc}).
Although 
we only have limited data points at $\mu < 2$, 
it is clear that there is no peak which grows
as the volume becomes larger.  
That is, there is no indication of the first order transition.

We need some care to interpret this result.
As the volume becomes larger with fixed fermion mass, 
the system moves away from the $\epsilon$-regime so that we need to use small 
enough fermion mass to stay in the $\epsilon$-regime.
From the smallest mass plot ($am=0.003$, top panel in Fig.~\ref{fig:susc}), 
the larger ($8^3\times 8$) volume gives the larger susceptibility.
This volume dependence might imply the first order transition, but 
since no peak structure is observed, our data cannot be used to draw this conclusion.
This is different from SU(3) case, where first order bulk phase transitions
related to the $S^4$ symmetry for staggered fermion exist \cite{Cheng:2011ic}.

If the transition is not first order, it 
can be either a crossover or a second order phase transition.   
The latter case implies possible existence of a non-trivial continuum limit in
the $\epsilon$-regime.
Since we can discuss only the broken side of the phase transition with our analysis,
independent estimations of the chiral susceptibility with
different method such as direct lattice calculation are required
to give a conclusive result.

\section{Conclusions}
\label{sec:conclusions}

We revised the chiral condensate of $n_f=8$ naive staggered system
in SU(2) fundamental representation 
by fitting the smallest Dirac eigenvalue with symplectic chiral RMT.  
We used a new numerical estimation 
of the distribution of individual eigenvalues of the RMT.
As previously reported, at strong coupling, we observe a bulk phase in which 
chiral symmetry is broken.  
The weak coupling side is the symmetric phase so that it is consistent
with the scenario that the theory is in the conformal window.
By further using RMT, we also have estimated
chiral susceptibility.  Although data in the $\epsilon$-regime,
in which our estimation is justified, is limited, we have not observed
any peak which grows with the volume in the susceptibility.
This implies that the bulk transition is not of first order.

\subsection*{Acknowledgments}

We thank C.~Y.~H.~Huang and K.~Ogawa for their contributions in generating configurations. 
I.K. is supported by part by
MEXT as ``Priority Issue 9 to be Tackled by Using Post-K Computer'' (Elucidation of the
Fundamental Laws and Evolution of the Universe) and JICFuS.
C.J.D.L acknowledges research grant 105-2628-M009-003-MY4
from Taiwanese MoST.

\end{document}